\documentclass[journal]{IEEEtran}
\usepackage{amsmath, amssymb, amsfonts, bm}
\usepackage{xcolor}
\usepackage{amsthm}
\usepackage{amsmath}
\usepackage{amssymb}
\usepackage{graphicx}
\usepackage{hyperref}    
\usepackage{enumitem}
\theoremstyle{remark}    

\theoremstyle{lemma}    

\theoremstyle{example}    

\newtheoremstyle{myproof}
  {}{}
  {\normalfont}          
  {}{\itshape}{.}
  { }{}
\theoremstyle{myproof}

\begin{document}
\title{Power Amplifier-aware Power Allocation for Noise-limited and Distortion-limited Regimes}
\author{\IEEEauthorblockN{Achref Tellili, Nathaniel Paul Epperson, Mohamed Akrout}\\
\IEEEauthorblockA{\textit{EECS Department}, \textit{University of Tennessee}, Knoxville, TN, USA,\\
\{atellili,\,neppers2\}@vols.utk.edu, makrout@utk.edu}
}

\maketitle

\begin{abstract}
The conventional power allocation strategy via water-filling relies on the premise that the power amplifier (PA) operates sufficiently below saturation such that a linear RF chain model holds. This work integrates the PA nonlinearity directly into the power allocation formulation, thereby removing the linearity assumption altogether and enabling operation in regimes where distortion noise is non-negligible. Leveraging the Bussgang theorem, we establish a statistical linearization of the PA's hard-limiting model to characterize the trade-off between signal gain and power-dependent distortion. We propose a projected gradient descent algorithm that optimizes power allocation while identifying an optimal spatial back-off strategy. We also derive a closed-form thermal noise variance threshold that separates the noise-limited and distortion-limited operating regimes as a function of the distortion noise variance and the channel Frobenius norm. Numerical simulations validate that our amplifier-aware strategy provides significant capacity gains in the saturation regime compared to standard water-filling.
\end{abstract}

\section{Introduction}
\subsection{Motivation and Prior Work}
The practical deployment of communication systems is inherently bounded by the physical constraints of the radio-frequency (RF) front-end, particularly the power amplifier (PA) \cite{holma2011lte,schenk2008rf}. As spectral efficiency and transmit power requirements increase for different applications, the intrinsic nonlinearities of the PA emerge as a dominant bottleneck. From a communications perspective, these nonlinear effects manifest as distortion noise and create a critical saturation on the achievable rate of the communication link \cite{bjornson2014massive}. Unlike thermal noise which is statistically stationary, additive, and independent of the transmit signal power, the distortion noise arising from PA nonlinearities is signal-dependent since its power scales with the transmit power, typically increasing more rapidly than the desired signal component under high back-off operation. For this reason, the distortion noise creates a fundamental power-dependent ceiling on the achievable rate, such that beyond a certain transmit power, further increases yield diminishing or even negative gains on capacity.

Standard communication theory often relies on the assumption of a linear RF chain, where the transmit signal is perfectly amplified within the available power budget \cite{heath2018foundations}. Under this idealized assumption, the water-filling (WF) algorithm is the canonical solution for power allocation, distributing power to the strongest eigen channels of the channel matrix $\mathbf{H}$. While mathematically elegant, this approach ignores the reality of gain compression and peak-to-average power ratio issues. In practical systems, PAs exhibit a nonlinear response as the input voltage approaches the saturation threshold $V_{\text{CC}}$, inducing signal distortion and out-of-band emissions. Prior work has utilized the Bussgang theorem \cite{bussgang1952crosscorrelation} to linearize these effects, modeling the PA output as a combination of a compressed coherent signal and an uncorrelated distortion noise term and an closed-form expression of the ergodic capacity has been characterized \cite{fozooni2015performance}. The optimal operating input back-off of the PA was derived in \cite{mezghani2010circuit} as a function of the path-loss (i.e. the communication distance). Another line of work devised a compensation scheme for the PA nonlinearity that defines the constellation and decision regions of the distorted transmitted signal in advance \cite{qi2010analysis}.

\subsection{Contributions}
In this paper, we revisit the power allocation problem under the PA non-linearity. Our work unfolds through the following contributions:
\begin{itemize}[leftmargin=*]
    \item After establishing the Bussgang decomposition of the PA output for hard-limiting PA models, we propose a projected gradient descent algorithm that identifies the optimal balance between coherent signal gain and distortion noise. This optimization reveals a spatial back-off strategy, where the system intentionally reduces power on high-gain eigen channels to prevent them from dominating the aggregate distortion noise.
    \item We derive a closed-form thermal noise variance threshold that separates the noise-limited and distortion-limited operating regimes. This threshold is not only a property of the PA but also tied to the spatial properties of the communication channel $\mathbf{H}$, namely its Frobenius norm.
    \item We validate through simulations how the proposed amplifier-aware water-filling strategy provides significant capacity gains (exceeding 100\% in deep saturation) compared to conventional water-filling.
\end{itemize}

\section{Statistical Linearization of the Power Amplifier's Behavior via Bussgang's Theorem}

To characterize the impact of power amplifier (PA) nonlinearities on communication performance, we employ the Bussgang theorem \cite{bussgang1952crosscorrelation}. This theorem is particularly powerful for analyzing memoryless nonlinearities when the input signal follows a Gaussian distribution. This stochastic characterization is key in facilitating information-theoretic analysis and serves as a canonical model for signals employing high-order modulation schemes and multi-carrier waveforms such as Orthogonal Frequency Division Multiplexing (OFDM) \cite{banelli2002theoretical}.

\subsection{The Bussgang Decomposition}

Let the transmit signal be a circularly-symmetric complex Gaussian random variable, denoted by $v_{\text{T}} \sim \mathcal{CN}(0, P_{\text{T}})$, where $P_{\text{T}}$ represents the average transmit power. When passed through a nonlinear operator $g(\cdot)$ representing the PA, the output $v_{\text{T},\text{PA}} = g(v_{\text{T}})$ is no longer Gaussian. However, the Bussgang theorem allows us to decompose this output into a coherent signal component and a distortion term:
\begin{equation}
    v_{\text{T,PA}} = g(v_{\text{T}}) = \alpha \,v_{\text{T}} + \eta_{\text{PA}},
\end{equation}
where $\alpha$ is the equivalent linear gain and $\eta_{\text{PA}}$ is the distortion noise. A fundamental property of this decomposition is that the distortion $\eta_{\text{PA}}$ is statistically uncorrelated with the input signal $v_{\text{T}}$, i.e., $\mathbb{E}\{v_{\text{T}}^* \,\eta_{\text{PA}}\} = 0$.

\noindent The complex gain $\alpha$ represents the effective amplification experienced by the signal after accounting for saturation. For a memoryless nonlinearity, $\alpha$ is defined by the cross-correlation between the input and output:
\begin{equation}\label{eq:alpha-def}
    \alpha = \frac{\mathbb{E}\{v_{\text{T,PA}} v_{\text{T}}^*\}}{\mathbb{E}\{|v_{\text{T}}|^2\}} = \frac{\mathbb{E}\{g(v_{\text{T}})\, v_{\text{T}}^*\}}{P_{\text{T}}}.
\end{equation}

Since $v_{\text{T}}$ is a complex Gaussian variable with variance $P_{\text{T}}$, we can represent it in polar coordinates as $v_{\text{T}} = r e^{j\phi}$. Here, $r$ is the magnitude which follows a Rayleigh distribution $f_R(r) = \frac{2r}{P_{\text{T}}} \exp\big(-r^2/P_{\text{T}}\big)$, while $\phi$ is the phase being uniformly distributed over $[0, 2\pi)$. We then write  $\mathbb{E}\{g(v_{\text{T}}) v_{\text{T}}^*\}$ as a double integral over the magnitude and phase
\begin{equation}\label{eq:expectation1}
    \mathbb{E}\{g(v_{\text{T}}) v_{\text{T}}^*\} = \int_{0}^{\infty} \int_{0}^{2\pi} g(r e^{j\phi}) \cdot (r e^{-j\phi}) \cdot f_{R,\Phi}(r, \phi) \, d\phi \, dr.
\end{equation}

\subsection{Bussgang decomposition for hard-limiting power amplifiers}

For a memoryless nonlinearity like a power amplifier, the transformation only affects the magnitude or adds a phase shift that depends only on the magnitude, i.e., $g(v_{\text{T}}) = g(r)e^{j\phi}$. Hence, (\ref{eq:expectation1}) can be expressed as
\begin{equation}
    \begin{aligned}
        \mathbb{E}\{g(v_{\text{T}}) v_{\text{T}}^*\} &= \int_{0}^{\infty} \int_{0}^{2\pi} g(r)\,e^{j\phi} \cdot (r \,e^{-j\phi}) \cdot f_R(r) \frac{1}{2\pi} \, d\phi \, dr\\
        &=\int_{0}^{\infty} g(r) \cdot r \cdot f_R(r) \, dr,
    \end{aligned}
\end{equation}
according to which $\alpha$ given in (\ref{eq:alpha-def}) becomes
\begin{equation}\label{eq:alpha-def}
    \alpha = \frac{1}{P_{\text{T}}}\int_{0}^{\infty} g(r) \cdot r \cdot f_R(r) \, dr.
\end{equation}

We consider an ideal hard-limiting model (a.k.a., soft-limiter) where the amplifier behaves linearly until the output voltage reaches the supply rails $V_{\text{CC}}$. Its output magnitude $g(r)$ is defined as:
\begin{equation}\label{eq:gr}
    g(r) = 
    \begin{cases} 
    G \cdot r, & r < \frac{V_{\text{CC}}}{G} \\
    V_{\text{CC}}, & r \geq \frac{V_{\text{CC}}}{G}
    \end{cases}
\end{equation}
where $G$ is the small-signal gain. We define the clipping threshold as $r_{\text{th}} = V_{\text{CC}}/G$.

\subsubsection{Derivation of the equivalent linear gain $\alpha$}

Substituting the definition of $g(r)$ in (\ref{eq:gr}) into the equivalent gain $\alpha$ established in (\ref{eq:alpha-def}) and and splitting the integral at the threshold $r_{\text{th}}$ yields
\begin{equation}
    \alpha = \frac{2}{P_{\text{T}}^2} \left( \int_{0}^{r_{\text{th}}} G \,r^3 e^{-r^2/P_{\text{T}}} \, dr + \int_{r_{\text{th}}}^{\infty} V_{\text{CC}} \,r^2 e^{-r^2/P_{\text{T}}} \, dr \right).
\end{equation}

\noindent The first integral represents the contribution of the linear regime, while the second captures the contribution of the saturation regime. after defining $k \triangleq V_{\text{CC}} / (G\sqrt{2 P_{\text{T}}})$ and by applying integration by parts and evaluating the limits, the expression simplifies to:
\begin{equation}
    \alpha(P_{\text{T}}) = G \cdot \text{erf}\left( \frac{V_{\text{CC}}}{G\sqrt{2 P_{\text{T}}}} \right),
\end{equation}
with $\textrm{erf}(\cdot)$ being the error function. This result shows that $\alpha$ is bounded by $[0, G]$. In the low-power regime ($P_{\text{T}} \to 0$), the error function approaches unity and $\alpha \approx G$. Conversely, as $P_{\text{T}} \to \infty$, the gain undergoes compression and vanishes, representing total signal saturation.

\subsubsection{Derivation of the distortion variance $\sigma_{\eta}^2$}

The distortion term $\eta_{\text{PA}}$ represents the wasted power associated to the portion of the signal lopped off by the clipping process. By the Bussgang theorem, the output power $P_{\text{out}}$ can be decomposed into coherent and uncorrelated components. The distortion variance $\sigma_{\eta}^2$ is thus the residual power:
\begin{equation}
    \sigma_{\eta}^2(P_{\text{T}}) = P_{\text{out}} - |\alpha|^2 P_{\text{T}}.
\end{equation}
The variance $\sigma_{\eta}^2$ is negligible in the linear regime but grows rapidly once $P_{\text{T}}$ exceeds the saturation threshold $(V_{\text{CC}}/G)^2$. This term acts as a self-generated interference floor that ultimately limits the signal-to-noise and distortion ratio (SNDR) and the resulting channel capacity. To find the variance $\sigma_{\eta}^2$, we must first determine the total output power $P_{\text{out}}$ of the nonlinear PA.

\noindent For a complex Gaussian input, the total output power is calculated by integrating the squared magnitude of the transfer function $g(r)$ over the Rayleigh distribution:
\begin{equation}
    P_{\text{out}} = \mathbb{E}\{|g(v_{\text{T}})|^2\} = \int_{0}^{\infty} |g(r)|^2 \frac{2r}{P_{\text{T}}} e^{-r^2/P_{\text{T}}} \, dr.
\end{equation}
Substituting the hard-limiting PA characteristic, the integral is split into a linear component and a saturated component:
\begin{equation}
    P_{\text{out}} = \frac{2}{P_{\text{T}}} \left[ \int_{0}^{r_{\text{th}}} G^2 r^3 e^{-r^2/P_{\text{T}}} \, dr + \int_{r_{\text{th}}}^{\infty} V_{\text{CC}}^2 r e^{-r^2/P_{\text{T}}} \, dr \right].
\end{equation}
Evaluating these integrals, we obtain:
\begin{equation}
    P_{\text{out}} = G^2 \left( P_{\text{T}} \,\text{erf}(k) - \frac{2 V_{\text{CC}} \sqrt{P_{\text{T}}}}{G \sqrt{2\pi}} e^{-k^2} \right) + V_{\text{CC}}^2 (1 - \text{erf}(k)).
\end{equation}

\begin{figure}[h!]
\vspace{-0.5cm}
\begin{minipage}[b]{0.9\linewidth}
  \centering
  \centerline{\includegraphics[scale=0.25]{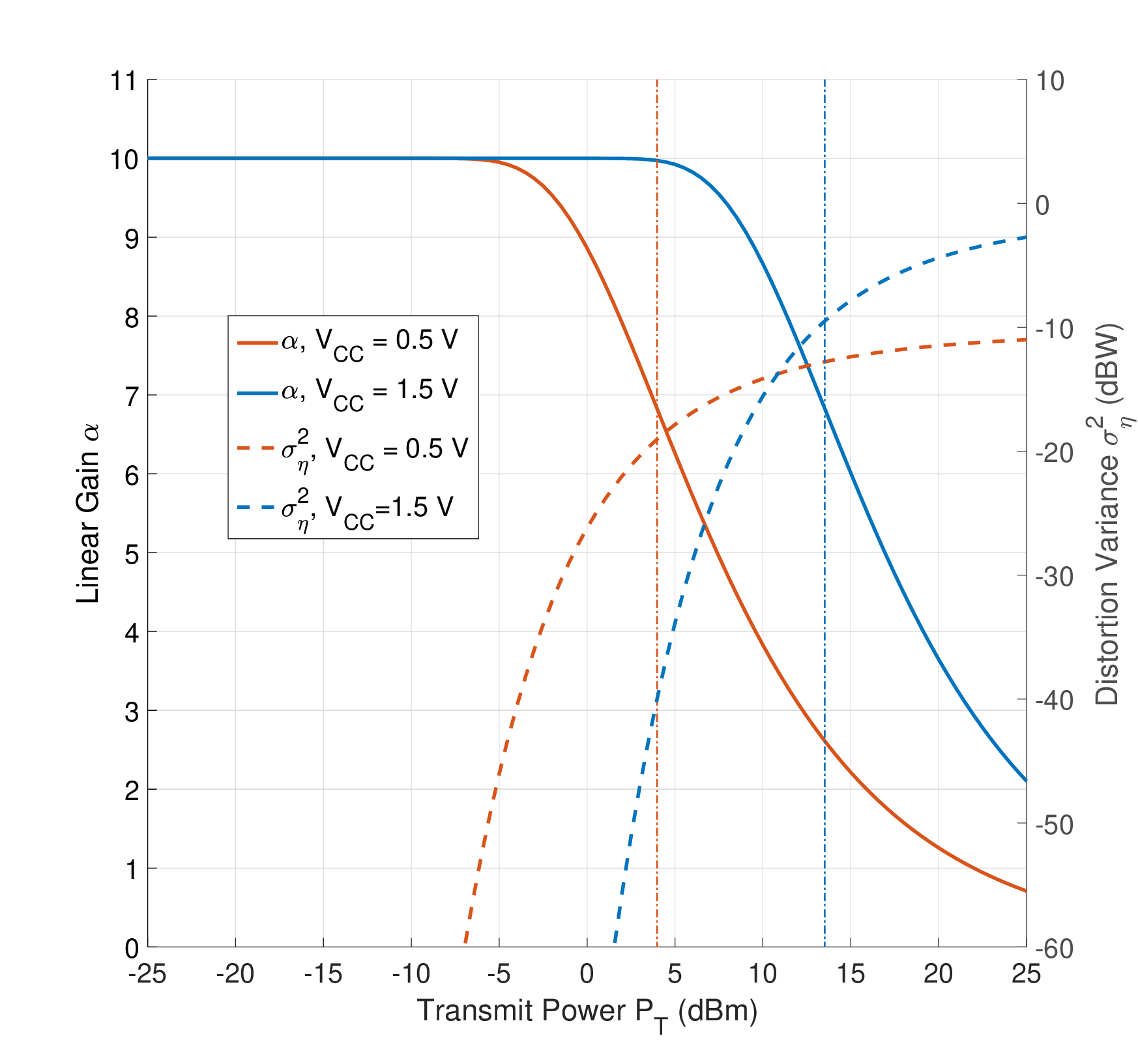}}
\end{minipage}
\vspace{-0.2cm}
\caption{Linear gain $\alpha$ and distortion variance $\sigma_{\eta}^2$ as a function of transmit power $P_{\text{T}}$ for varied supply voltages $V_{\text{CC}}$. The vertical dotted lines indicate the theoretical saturation power thresholds $P_{\text{sat}} = (V_{\text{CC}}/G)^2$ between the linear and nonlinear operational regimes.}
\label{fig:alpha-eta}
\end{figure}

The impact of the power amplifier's supply voltage $V_{\text{CC}}$ on the signal integrity is illustrated in Fig. \ref{fig:alpha-eta}. We observe that for low transmit powers, the equivalent gain $\alpha$ remains constant at the small-signal gain $G=10$, while the distortion variance $\sigma_{\eta}^2$ is negligible, representing a strictly linear operational regime. However, as the input power approaches the saturation threshold $P_{\text{sat}} = (V_{\text{CC}}/G)^2$ (displayed by the vertical dotted lines), the gain begins to undergo compression. Crucially, a lower $V_{\text{CC}}$ of $0.5$\,V causes the amplifier to hit the voltage rails significantly earlier than the $1.5$\,V case, leading to a precipitous drop in $\alpha$ and a simultaneous $20$\,dB increase in the distortion floor at moderate power levels (e.g., $0$\,dBm). This behavior highlights a fundamental shift in the analysis of the communication RF chain: unlike the additive white Gaussian noise (AWGN) model where the noise floor is independent of the transmit signal, the distortion variance $\sigma_{\eta}^2$ is a strictly power-dependent quantity. As a result, the classical definition of the signal-to-noise ratio (SNR) fails to capture the true link quality in the saturation regime, as any increase in $P_{\text{T}}$ simultaneously elevates the self-generated interference floor. This coupling implies that standard power allocation strategies, such as water-filling, which assume a static noise power, are no longer optimal and may even lead to a reduction in capacity. For this reason, a new power allocation method must be devised to balance the coherent gain $\alpha$ against the power-dependent distortion $\sigma_{\eta}^2$ to prevent capacity collapse.

\section{Capacity Analysis and Noise-Distortion Regimes}

In a MIMO transmitter with $N_{\text{T}}$ antennas, each RF chain is equipped with an independent power amplifier. To model the aggregate effect of these nonlinearities, the scalar Bussgang decomposition must be extended to a vectorized form. Let $\mathbf{v}_{\text{T}} \in \mathbb{C}^{N_{\text{T}} \times 1}$ be the vector of Gaussian transmit signals. The output of the nonlinear array is given by:
\begin{equation}\label{eq:sysmodel-vPA}
    \mathbf{v}_{\text{T,PA}} = \mathbf{A} \mathbf{v}_{\text{T}} + \boldsymbol{\eta}_{\text{PA}},
\end{equation}
where $\mathbf{A} = \text{diag}(\alpha_1, \dots, \alpha_{N_{\text{T}}})$ is the diagonal gain matrix and $\boldsymbol{\eta}_{\text{PA}} = [\eta_1, \eta_2, \dots, \eta_{N_{\text{T}}}]^\top \in \mathbb{C}^{N_{\text{T}} \times 1}$ is the aggregate distortion noise vector. Because transmit RF chains are assumed independent, $\boldsymbol{\eta}_{\text{PA}}$ is a zero-mean complex Gaussian random vector with a diagonal covariance matrix
\begin{equation}
    \mathbf{R}_{\eta} = \mathbb{E}\{\boldsymbol{\eta}_{\text{PA}} \,\boldsymbol{\eta}_{\text{PA}}^H\} = \text{diag}(\sigma_{\eta, 1}^2, \sigma_{\eta, 2}^2, \dots, \sigma_{\eta, N_{\text{T}}}^2),
\end{equation}
where each diagonal entry $\sigma_{\eta, i}^2$ is the power-dependent distortion variance specifically associated with the $i$-th power amplifier's operating point owing to the Bussgang theorem. Each entry $\alpha_i$ is a function of the local transmit power $P_i$ at the $i$-th antenna. The distortion vector $\boldsymbol{\eta}_{\text{PA}}$ follows a zero-mean distribution with a diagonal covariance matrix $\mathbf{R}_{\eta} = \text{diag}(\sigma_{\eta, 1}^2, \dots, \sigma_{\eta, N_{\text{T}}}^2)$, assuming the amplifiers are statistically independent.

After letting $\mathbf{p} = [P_1, P_2, \dots, P_{N_{\text{T}}}]^T$ denote the power allocation vector where $P_i = \mathbb{E}\{|v_{\text{T},i}|^2\}$ is the average transmit power allocated to the $i$-th antenna, we use the output PA voltage in (\ref{eq:sysmodel-vPA}) to write the received signal $\mathbf{v}_{\text{L}}$ at the receive load is expressed as
\begin{equation}
    \mathbf{v}_{\text{L}}(\mathbf{p}) = \mathbf{H}\,\mathbf{v}_{\text{T}}(\mathbf{p}) + \mathbf{n} =  \underbrace{\mathbf{H} \,\mathbf{A}(\mathbf{p})}_{\triangleq\,\mathbf{H}_{\textrm{eff}}(\mathbf{p})} \,\mathbf{v}_{\text{T}} + \underbrace{\mathbf{H} \,\boldsymbol{\eta}_{\text{PA}}(\mathbf{p}) + \mathbf{n}}_{_{\triangleq\,\mathbf{n}_{\textrm{eff}}(\mathbf{p})}}.
\end{equation}

\noindent The effective channel, $\mathbf{H}_{\textrm{eff}}(\mathbf{p})$ illustrates that the spatial properties of the physical channel are now weighted by the power-dependent diagonal gain matrix $\mathbf{A}(\mathbf{p})$, where each entry $\alpha_i(P_i)$ accounts for the specific compression state of the $i$-th transmit chain. Moreover, the effective noise vector $\mathbf{n}_{\textrm{eff}}$ reveals that the receiver is subject not only to additive thermal noise but also to transmitter-side distortion that has been spatially filtered by the channel matrix $\mathbf{H}$. Unlike classical linear models where the PA is not incorporated, the total effective noise here is colored by the channel matrix $\mathbf{H}$, meaning the transmitter's nonlinearity can contaminate the entire spatial subspace at the receiver.

Under the assumption that the thermal noise and PA distortion are statistically independent, the covariance matrix of the effective noise is expressed as:
\begin{equation}\label{eq:neff-covariance}
    \mathbf{R}_{\mathbf{n}_{\text{eff}}}(\mathbf{p}) = \mathbb{E}\Big\{ \mathbf{n}_{\text{eff}}(\mathbf{p}) \,\mathbf{n}_{\text{eff}}(\mathbf{p})^{\mathsf{H}} \Big\} = \mathbf{H} \,\mathbf{R}_{\eta}(\mathbf{p}) \,\mathbf{H}^H + \mathbf{R}_n,
\end{equation}
where $\mathbf{R}_{\eta}(\mathbf{p}) = \text{diag}\left(\sigma_{\eta, 1}(P_1)^2, \sigma_{\eta, 2}(P_2)^2, \dots, \sigma_{\eta, N_{\text{T}}}(P_{N_{\text{T}}})^2\right)$ represents the signal-dependent distortion variances and $\mathbf{R}_n = \sigma_n^2 \mathbf{I}_{N_{\text{R}}}$ represents the covariance of the additive thermal noise at the $N_{\text{R}}$ receive antennas. This formulation underscores a critical departure from classical MIMO theory: the noise covariance is no longer a static identity matrix but a dynamic function of the transmit power allocation, necessitating a new power allocation optimization to maintain link integrity in the saturation regime.

\subsection{Capacity under thermal and distortion noise}
Assuming a zero-mean Gaussian signaling input $\mathbf{v}_{\text{T}} \sim \mathcal{CN}(\mathbf{0}, \mathbf{R}_v(\mathbf{p}))$ where $\mathbf{R}_v(\mathbf{p}) = \text{diag}(P_1, \dots, P_{N_{\text{T}}})$, the capacity $C$ (in bits/s/Hz) is expressed as \cite{heath2018foundations}:

\begin{equation}\label{eq:capacity_det}
    \begin{aligned}
    C(\mathbf{p}) &= \log_2 \det \left( \mathbf{I}_{N_{\text{R}}} + \mathbf{R}_{\mathbf{n}_{\text{eff}}}^{-1}(\mathbf{p}) \mathbf{H}_{\text{eff}}(\mathbf{p}) \mathbf{R}_v(\mathbf{p}) \mathbf{H}_{\text{eff}}(\mathbf{p})^{\mathsf{H}} \right)\\
    &=\log_2 \det \Big( \mathbf{I}_{N_{\text{R}}} + \left( \mathbf{H} \mathbf{R}_{\eta}(\mathbf{p}) \mathbf{H}^{\mathsf{H}} + \sigma_n^2 \mathbf{I}_{N_{\text{R}}} \right)^{-1}\\
    &\hspace{3cm}\times\,\mathbf{H} \mathbf{A}(\mathbf{p}) \mathbf{R}_v(\mathbf{p}) \mathbf{A}(\mathbf{p})^{\mathsf{H}} \mathbf{H}^{\mathsf{H}} \Big).
    \end{aligned}
\end{equation}

\noindent We seek the optimal power allocation vector $\mathbf{p}$ under a total transmit power constraint $P_{\text{total}}$. The optimization problem is formulated as follows:
\begin{equation}\label{eq:problem-formulation}
    \begin{aligned}
        \mathcal{P}: \max_{\mathbf{p}} \quad & C(\mathbf{p}) \\
        \text{such that} \quad & \sum_{i=1}^{N_{\text{T}}} P_i \leq P_{\text{total}}, \quad P_i \geq 0, \forall i,
    \end{aligned}
\end{equation}

\noindent To solve (\ref{eq:problem-formulation}), we analyze the Lagrangian $\mathcal{L}(\mathbf{p}, \lambda, \boldsymbol{\mu}) = f(\mathbf{p}) - \lambda (\sum P_i - P_{\text{total}}) + \sum \mu_i P_i$. The stationarity condition with respect to $P_i$ is given by:
\begin{equation}\label{eq:kkt_grad}
    \frac{\partial C(\mathbf{p})}{\partial P_i} = \text{tr} \left( \mathbf{K}^{-1} \frac{\partial \mathbf{Q}}{\partial P_i} \right) - \text{tr} \left( \mathbf{K}^{-1} \mathbf{Q} \mathbf{K}^{-1} \frac{\partial \mathbf{R}_{\mathbf{n}_{\text{eff}}}}{\partial P_i} \right) = \lambda,
\end{equation}
where $\mathbf{Q}(\mathbf{p}) = \mathbf{H} \mathbf{A}(\mathbf{p}) \mathbf{R}_v(\mathbf{p}) \mathbf{A}(\mathbf{p})^{\mathsf{H}} \mathbf{H}^{\mathsf{H}}$ and $\mathbf{K} = \mathbf{R}_{\mathbf{n}_{\text{eff}}}(\mathbf{p}) + \mathbf{Q}(\mathbf{p})$. When the effect of the PA is ignored, $\frac{\partial \mathbf{R}_{\mathbf{n}_{\text{eff}}}}{\partial P_i} = \mathbf{0}$, and power is allocated solely based on the first term of (\ref{eq:kkt_grad}). When the PA is incorporated into the analysis, (\ref{eq:kkt_grad}) introduces a negative penalty term proportional to the distortion gradient $\frac{\partial \sigma_{\eta,i}^2}{\partial P_i}$. Since this distortion is spatially colored by $\mathbf{H}$, the power optimizer must reduce $P_i$ for antennas that contribute significantly to the aggregated distortion noise level floor \textit{even if they possess a strong channel}.

\subsection{Power allocation under power amplifier non-linearity}
Since the optimization problem $\mathcal{P}$ involves a non-convex objective, we propose a projected gradient descent (PGD) algorithm to find the optimal power allocation strategy. Unlike standard water-filling,which admits a closed-form solution via Lagrange multipliers, the coupling between $P_i$ and $\mathbf{R}_{\mathbf{n}_{\text{eff}}}$ requires an iterative approach. The gradient of the capacity with respect to the $i$-th antenna power $P_i$ is derived using the chain rule on the log-determinant function. The gradient is given by:
\begin{equation}
    \nabla_i f(\mathbf{p}) = \frac{1}{\ln 2} \left[ \text{tr} \left( \mathbf{K}^{-1} \frac{\partial \mathbf{Q}}{\partial P_i} \right) - \text{tr} \left( \mathbf{K}^{-1} \mathbf{Q} \mathbf{K}^{-1} \frac{\partial \mathbf{R}_{\mathbf{n}_{\text{eff}}}}{\partial P_i} \right) \right].
\end{equation}
The partial derivatives of the Bussgang parameters are:
\begin{subequations}
    \begin{align}
    \frac{\partial \alpha_i}{\partial P_i} &= -\frac{V_{\text{CC}} \exp(-k_i^2)}{P_i \sqrt{2\pi P_i}},\\
    \frac{\partial \sigma_{\eta,i}^2}{\partial P_i} &= \frac{\partial P_{\text{out},i}}{\partial P_i} - \left( 2\alpha_i P_i \frac{\partial \alpha_i}{\partial P_i} + \alpha_i^2 \right).
    \end{align}
\end{subequations}

\noindent After each gradient step, the updated power vector $\tilde{\mathbf{p}}$ may violate the total power constraint or the non-negativity constraint. We apply a projection operator $\Pi_{\mathcal{C}}(\cdot)$ onto the convex set $\mathcal{C} = \{ \mathbf{p} : \mathbf{1}^{\mathsf{T}}\mathbf{p} \leq P_{\text{total}}, P_i \geq 0 \}$. This is achieved at any iteration $t$ via a water-filling-like thresholding operation
\begin{equation}
    P_i^{(t)} = \max\Big(0, \widetilde{P}_i^{(t)} - \mu\Big),
\end{equation}
where $\mu$ is the dual variable chosen such that $\sum \max(0, \widetilde{P}_i^{(t)} - \mu) = P_{\text{total}}$. This iterative process ensures that the algorithm converges to a stationary point that respects both the power budget and the PA non-linearity. One can think of this power allocation algorithm within a dynamic control loop with variable attenuators and amplifiers adjusting their separate input voltage $V_{\textrm{CC}}$ as a function of the estimated channel.

\subsection{Noise-Distortion Transition Threshold}

To establish a rigorous boundary between the noise-limited regime and the distortion-limited regime. This boundary is defined as the point where the thermal noise power is equivalent to the aggregate distortion power at the receiver load. Recall the total effective noise covariance in (\ref{eq:neff-covariance}) defined as
\begin{equation}\label{eq:noise-covariance-explanation}
    \mathbf{R}_{\mathbf{n}_{\text{eff}}}(\mathbf{p}) = \underbrace{\mathbf{H} \,\mathbf{R}_{\eta}(\mathbf{p}) \,\mathbf{H}^{\mathsf{H}}}_{\textrm{distortion noise covariance}} + \underbrace{\mathbf{R}_n}_{\textrm{thermal noise covariance}}.
\end{equation}
Here, the term $\mathbf{H} \mathbf{R}_{\eta}(\mathbf{p}) \mathbf{H}^{\mathsf{H}}$ scales with power and overtake the thermal noise and causing a capacity collapse. Identifying the transition point allows us to determine when classical water-filling ceases to be effective and when amplifier-aware power allocation becomes critical. Toward this goal, we equate the traces of the thermal and distortion covariances matrices in (\ref{eq:noise-covariance-explanation}) as follows:
\begin{equation}\label{eq:threshold-derivation}
\begin{aligned}[b]
\text{Tr}\Big(\mathbf{R}_n\Big) &= \text{Tr}\Big(\mathbf{H} \,\mathbf{R}_{\eta}(\mathbf{p}) \,\mathbf{H}^{\mathsf{H}}\Big) \\
\iff N_{\text{R}} \,\sigma_{n}^2 &= \sigma_{\eta}^2(P_{\text{avg}}) \,\|\mathbf{H}\|_{\textrm{F}}^2,
\end{aligned}
\end{equation}
where the last equality is obtained assuming a reference uniform power allocation where $\mathbf{R}_{\eta} = \sigma_{\eta}^2(P_{\text{avg}}) \mathbf{I}_{N_{\text{T}}}$, and $\|\mathbf{H}\|_{\textrm{F}}$ is the Frobenius norm of the channel matrix. Finally, solving (\ref{eq:threshold-derivation}) for $\sigma_{n}^2$ yields the thermal noise variance threshold:
\begin{equation}\label{eq:threshold}
    \sigma_{n,\text{th}}^2 = \sigma_{\eta}^2(P_{\text{avg}}) \frac{\|\mathbf{H}\|_{\textrm{F}}^2}{N_{\text{R}}}.
\end{equation}
The threshold $\sigma_{n,\text{th}}^2$ in (\ref{eq:threshold}) implies two operating modes:
\begin{itemize}
    \item[$i)$] \textit{Noise-limited regime} when $\sigma_n^2 \gg \sigma_{n,\text{th}}^2$ when the capacity is dominated by thermal effects and classical water-filling is near-optimal.
    \item[$ii)$] \textit{Distortion-limited regime} when $\sigma_n^2 \ll \sigma_{n,\text{th}}^2$ : The capacity is limited by the distortion of the PA. In this regime, the noise is spatially colored by the channel $\mathbf{H}$, and amplifier-aware power optimization is mandatory to prevent capacity collapse.
\end{itemize}
The dependency of the thermal noise threshold $\sigma_{n,\text{th}}^2$ in (\ref{eq:threshold}) on the channel matrix $\mathbf{H}$ reveals that high-gain channels accelerate the onset of the distortion-limited regime. Specifically, the term $\|\mathbf{H}\|_{\textrm{F}}^2 / N_{\text{R}}$ represents the average spatial gain of the MIMO link. In high-SNR or near-field scenarios where the channel is exceptionally strong, the distortion induced by the PA is amplified just as effectively as the desired signal. This causes the distortion noise floor to rise well above the ambient thermal noise much earlier than it would in a fading or long-range far-field environment.

\section{Numerical Results and Discussions}

We assess the performance of the proposed amplifier-aware waterfilling algorithm compared to the standard waterfilling one. We let $N_{\textrm{T}} = N_{\textrm{R}} = 32$, $G=10$, and $V_{\textrm{CC}}=1\,V$.

\begin{figure}[h!]
\vspace{-0.4cm}
\begin{minipage}[b]{0.98\linewidth}
  \centering
  \centerline{\includegraphics[scale=0.28]{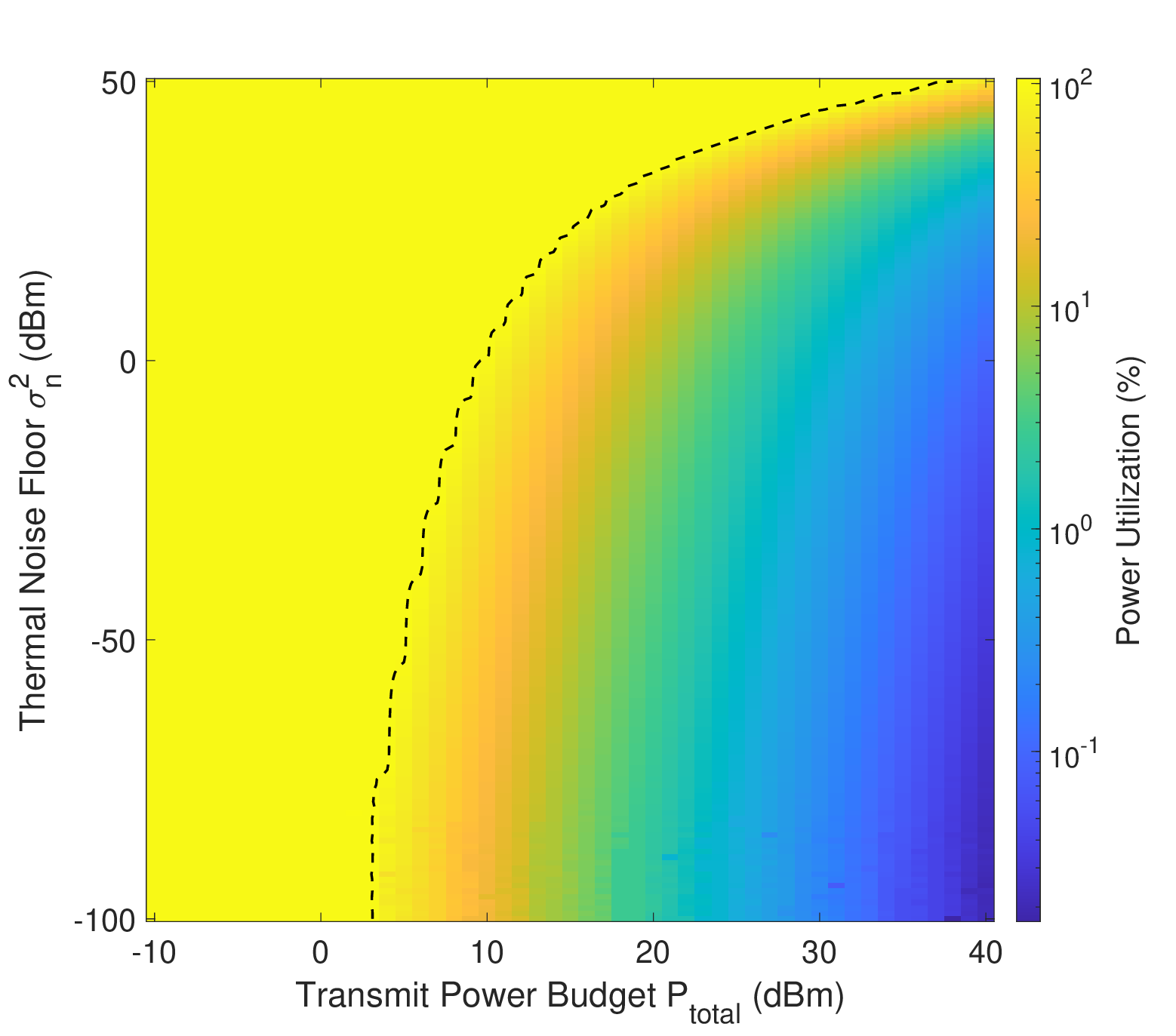}}
\end{minipage}
\vspace{-0.2cm}
\caption{Power utilization percentage versus transmit power budget $P_{\text{total}}$ (dBm) and thermal noise floor $\sigma_n^2$ (dBm). The dashed black line represents the theoretical saturation power threshold $P_{\text{sat}} = (V_{\text{CC}}/G)^2$.}
\label{fig:threshold-heatmap}
\end{figure}

Fig.~\ref{fig:threshold-heatmap} illustrates the power utilization percentage as a function of the transmit power budget as $100\cdot\Vert \mathbf{p} \Vert_1/P_{\textrm{total}}$. We observe that at low transmit power budgets, power utilization remains high across various noise levels. Above 10 dBm, power utilization decreases significantly in low-noise environments, reflecting the saturation effects of the PA.

\begin{figure}[h!]
\vspace{-0.4cm}
\begin{minipage}[b]{0.98\linewidth}
  \centering
  \centerline{\includegraphics[scale=0.28]{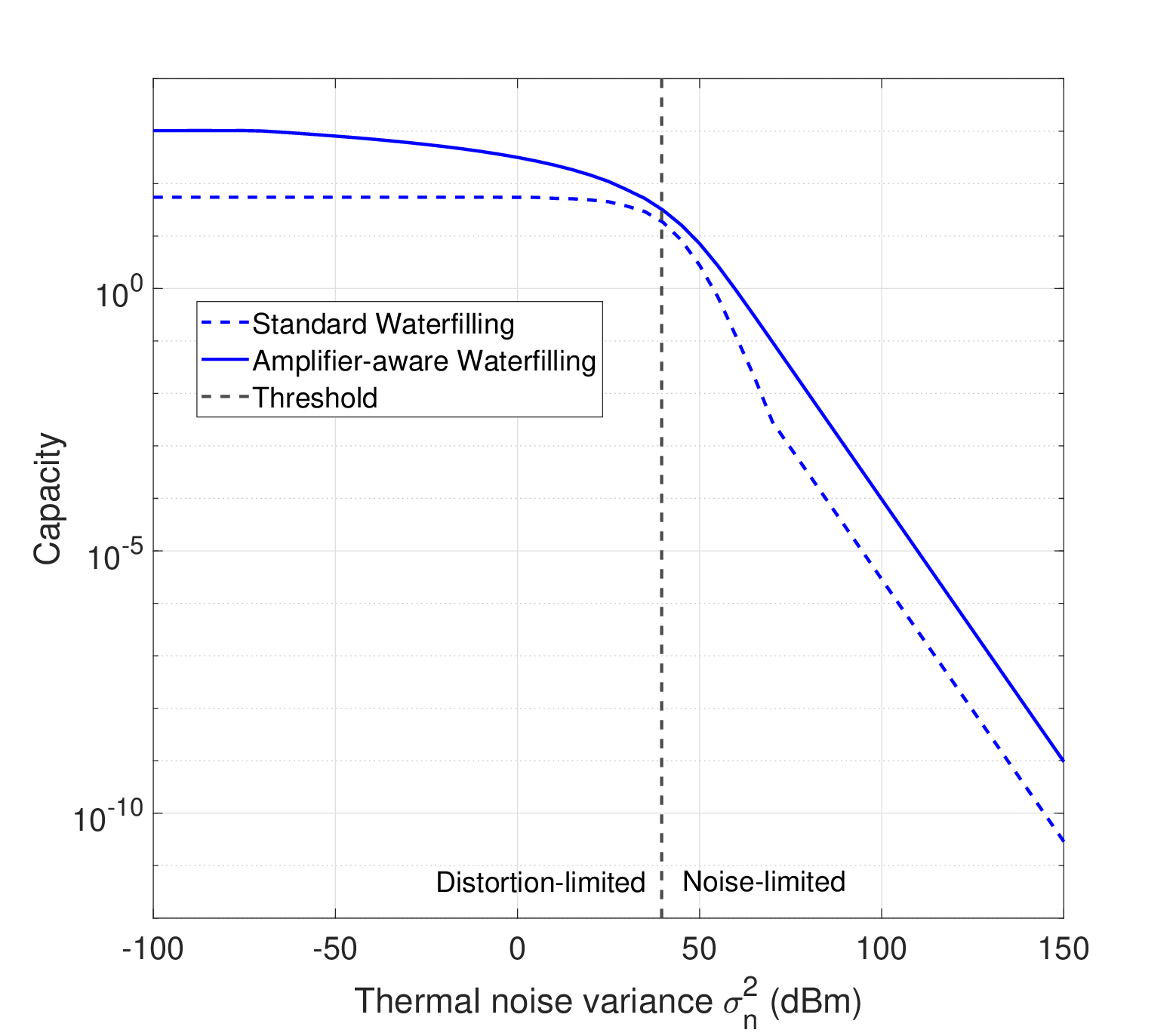}}
\end{minipage}
\vspace{-0.2cm}
\caption{Capacity versus thermal noise variance $\sigma_n^2$ (dBm) showing the transition from distortion-limited to noise-limited regimes.}
\label{fig:capacity-thermalnoise}
\end{figure}

The impact of thermal noise variance $\sigma_n^2$ on capacity is shown in Fig.~\ref{fig:capacity-thermalnoise}. The plot reveals two distinct operating regimes separated by the threshold established in (\ref{eq:threshold}): the \textit{distortion-limited} regime and the \textit{noise-limited} regime. In the distortion-limited regime (i.e., low $\sigma_n^2$), the amplifier-aware waterfilling outperforms standard waterfilling by accounting for the PA non-linearity. As the noise variance increases beyond the threshold (approx. 40 dBm), both methods converge as the system becomes noise-limited. While a thermal noise variance of $40$ dBm is significantly higher than standard communication noise floors, it can represents interference levels in specialized applications such as electronic warfare jamming or high-intensity solar radio flares captured by large-aperture systems.

\begin{figure}[h!]
\vspace{-0.4cm}
\begin{minipage}[b]{0.98\linewidth}
  \centering
  \centerline{\includegraphics[scale=0.28]{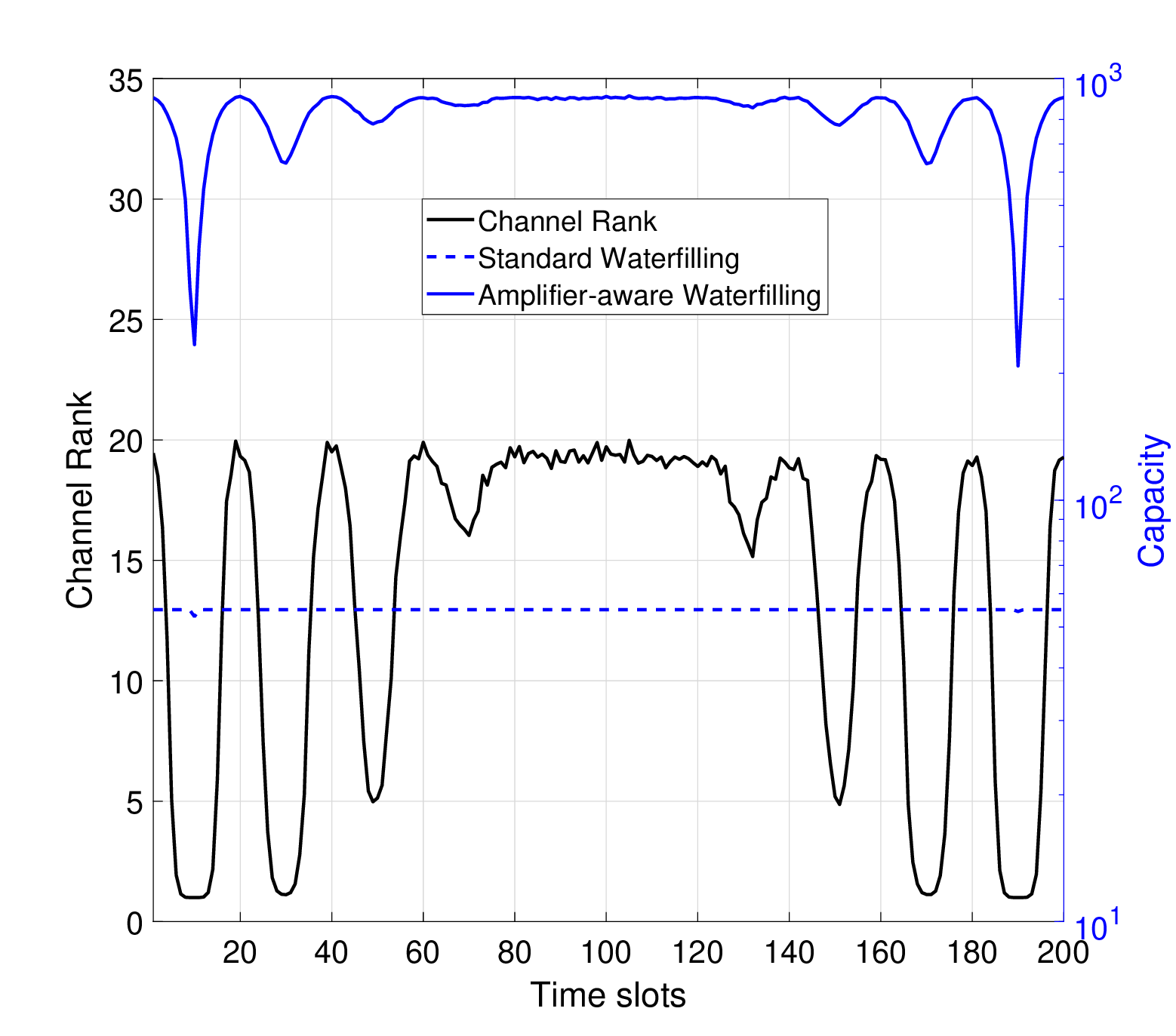}}
\end{minipage}
\vspace{-0.2cm}
\caption{Evolution of channel rank and capacity over 200 time slots for standard and amplifier-aware waterfilling with $P_{\text{total}} = 40\,\textrm{dBm}$ and $\sigma_n^2=10^{-9}$.}
\label{fig:waterfilling}
\vspace{-0.2cm}
\end{figure}

Finally, Fig.~\ref{fig:waterfilling} depicts the temporal behavior of the channel rank and the corresponding capacity over 200 time slots as we vary the channel from a Rayleigh fading model to multi-path one. While the channel rank fluctuates significantly between 0 and 20, the amplifier-aware waterfilling maintains a consistently higher capacity compared to the standard approach. This demonstrates the robustness of the proposed algorithm in tracking channel variations while staying within the optimal operating range of the power amplifiers.

\section{Conclusion}
This paper addresses the power amplifiers's saturation regime in multi-antenna systems, where excessive transmit power degrades capacity due to signal-dependent distortion. We derived a closed-form threshold for the thermal noise variance that identifies the transition between noise-limited and distortion-limited regimes. We also devised an amplifier-aware power allocation strategy using projected gradient descent, which outperforms conventional water-filling with significant capacity gains in the saturation regime.

\section*{Acknowledgement}
We thank Dr. Daniel Costinett and Dr. Benjamin J. Blalock for valuable discussions on the abstraction of power amplifier for saturation modeling.

\bibliographystyle{IEEEtran}
\bibliography{references.bib}    
\end{document}